\pdfoutput=1

\documentclass[
    ,final            
  ,pdftex                 
  ]
  {aipproc}

\layoutstyle{6x9}

\begin{document}

\title{Searching for Stopped Gluinos at CMS}



\author{Fedor Ratnikov for CMS Collaboration}{
  address={Institut f\"ur Experimentelle Kernphysik,
           Universit\"at Karlsruhe (TH),
           Postfach~6980,
           76128~Karlsruhe, Germany}
           ,altaddress={on leave from Institute for Theoretical and Experimental Physics, Moscow}
}

\begin{abstract}
We describe plans for a search for long-lived particles which will become
stopped by the CMS detector. We will look for the subsequent decay of these particles
during time intervals where there are no $pp$ collisions in CMS: during gaps between 
crossings in the LHC beam structure, and during inter-fill periods between the beam 
being dumped and re-injection. Such long living particles decays will be recorded 
with dedicated calorimeter triggers. 
For models predicting these particles, such as split-susy gluinos,
the large cross-section combined with good stopping power
of CMS, yields a significant number of triggerable decays. 
If LHC instantaneous luminosity approaches $10^{32} cm^{-2} s^{-1}$ in 2009-10, 
$5\sigma$ significance can be established
in a matter of days, since these decays occur on top of a negligible background.

Due to limited size, this paper concentrates on  main idea and expected results.  
More details are available in {\em https://twiki.cern.ch/twiki/bin/view/CMS/PhysicsResults}.
\end{abstract}

\maketitle


\label{sec:intro}

	There are a number of new physics scenarios which predict the existence of new heavy quasi-stable charged particles.  One such theoretical scenario is ``split supersymmetry"~\cite{ArkaniHamed:2004fb}.  As in more traditional supersymmetric models, in split SUSY copious gluino production is expected at the LHC via $gg \rightarrow \tilde{g}\tilde{g}$, with rates approaching 1 Hz (at the design luminosity) for the lightest gluino masses.  Unlike traditional SUSY, however, due to a very large mass splitting between the new scalars and new fermions, such gluinos can only decay through a highly virtual squark.  The lifetime of the gluino can thus be quite long; the gluino may well be stable on typical CMS~\cite{Adolphi:2008zzk} experimental timescales.  Existing experimental constraints on the value of this lifetime are weak~\cite{Arvanitaki:2005fa}.
	
	If long-lived gluinos are produced at CMS, they will hadronize into $\tilde{g}g, \tilde{g}q\bar{q}, \tilde{g}qqq$ states, which are  known as ``R-hadrons". Some of these gluino bound states will be charged, whilst others will be neutral.  Those, which are charged, will lose energy via ionization, as they traverse the CMS detector.  For low-$\beta$ R-hadrons, this energy loss will be sufficient to bring a significant fraction of the produced particles to rest inside the CMS detector volume~\cite{Arvanitaki:2005nq}.   These ``stopped" R-hadrons will decay seconds, days, or weeks later.  These decays will be out-of-time with respect to LHC collisions and may well occur at times when there are no collisions, e.g. beam gaps, or when there is no beam in the LHC machine, e.g. interfill periods.  The observation of such decays, when we expect  quiet detector conditions with only occasional cosmic rays,  
would be an unambiguous discovery of new physics.    
	
	We have devised and implemented a search strategy, that enables us to use CMS to detect such decays.

	Stopped particles  will decay in a location displaced by meters from the nominal interaction point of CMS, and asynchronously with LHC clock.  In order to study properties of these unusual events, and to learn how to trigger on and reconstruct them, we developed a custom simulation.  The simulation is factorized into three phases.  First we simulate gluino production, hadronization, and passage of a stable R-hadron through the CMS detector, and note at which point the particle has come to rest.   This process is repeated many times to obtain a distribution of stopping locations in the detector volume. Then we generate a particle at rest at that location, and simulate its decay. The full CMS simulation and data processing chain is applied to obtain the detector response to this stopped particle decay, and to obtain trigger and reconstruction signatures and efficiencies.   Finally, we simulate  the time of this decay relative to the LHC beam clock, taking into account particle lifetime, specific LHC beam structure, and expected LHC luminosity profile. 

\begin{figure}[th]
          \includegraphics[height=.198\textheight]{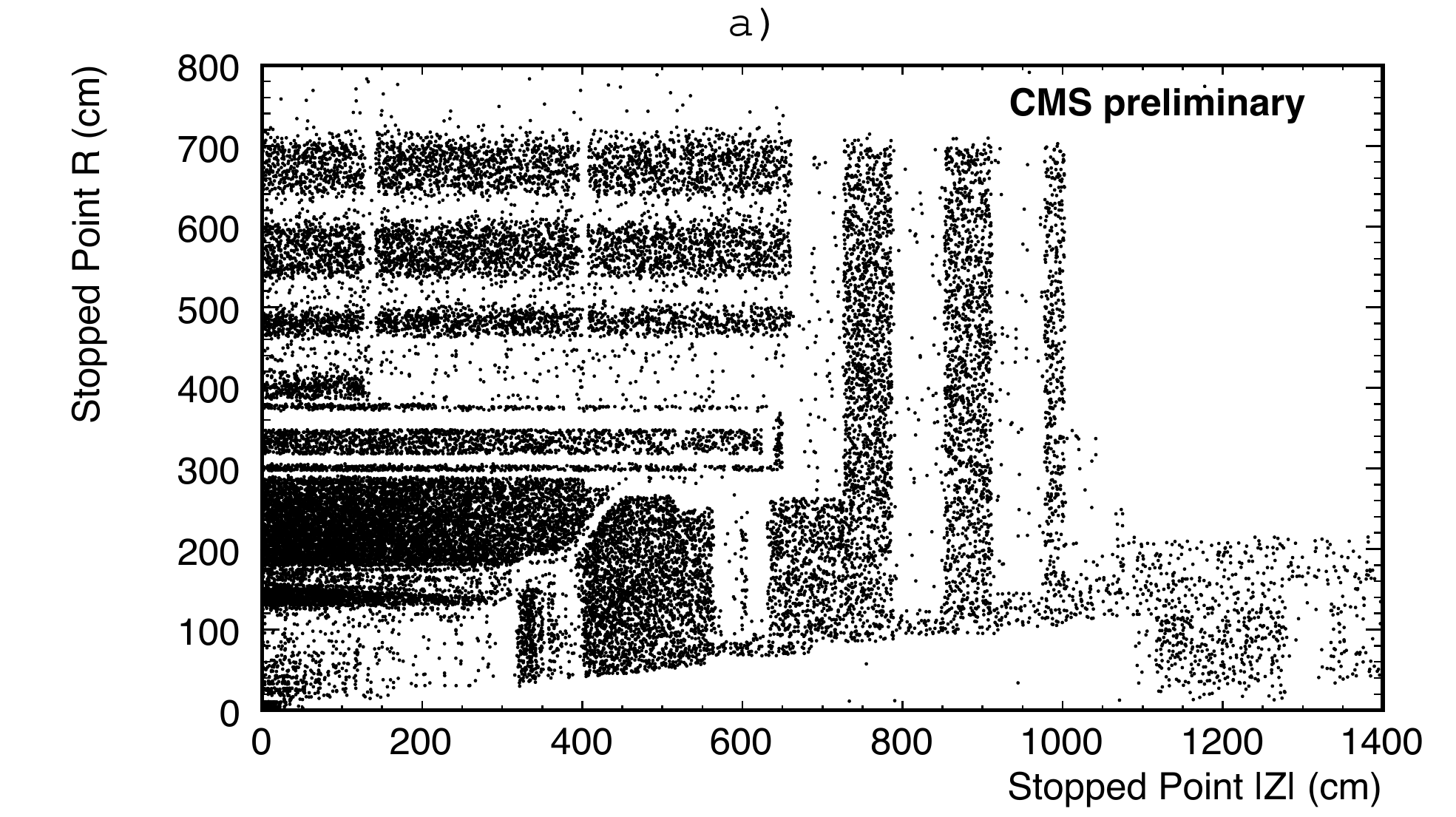}
          \includegraphics[height=.198\textheight]{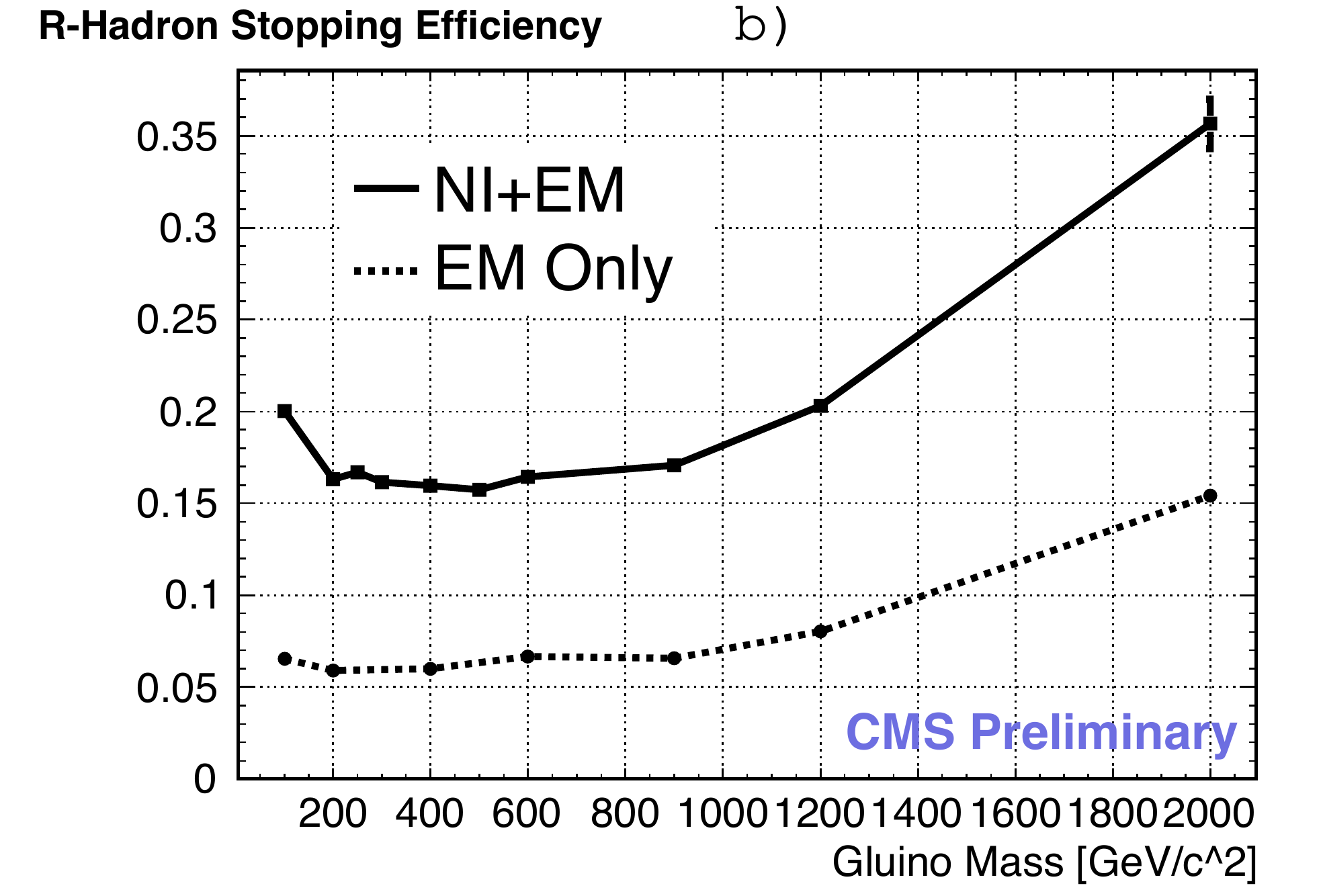}
       \caption{{\bf Left:} R-hadron stopping points for $m_{\tilde{g}}=300$ GeV, and $\sqrt{s} = 10$ TeV in the $R-|z|$ plane.
                {\bf Right:} Probability for the produced R-hadron to stop anywhere inside the CMS detector.
       for different gluino masses, and $\sqrt{s} = 10$ TeV.  The solid line shows stopping probability for both electromagnetic and nuclear interactions, while the dashed line shows that for electromagnetic interactions only.}
       \label{fig:stop}
\end{figure}

Figure~\ref{fig:stop} illustrates the distribution of stopping points. The material structure of the CMS detector is clearly seen in this distribution.
Figure~\ref{fig:stop} also presents the stopping probability as a function of gluino mass. The cloud model \cite{Mackeprang:2006gx}, used to describe nuclear interactions (NI) during propagation through the material, introduces an uncertainty in the stopping efficiency.  We therefore show the stopping probability due to the much more certain electromagnetic interactions (EM) alone alongside of the stopping efficiency due to the combined effects of EM+NI. 

A majority of particles come to rest in the calorimeters, most of them deposit significant energy in following decay. We designed and implemented specific calorimeter trigger to select such events.   
At L1 this trigger requires the normal 10 GeV jet, together with ``no collision'' condition.  We run this trigger when there is obviously no beam in the LHC, i.e. between fills and during shut-downs, as well during regular LHC luminosity periods.
The later extends lifetime sensitivity range from minutes to weeks for off-beam measurements, by range of lifetimes from $\mu$s to minutes for in-beam measurements, taking advantage of the numerous periods during the LHC orbit in which no $pp$ collisions occur in CMS.   These gaps have lengths of 0.2 $\mu s$, 1 $\mu s$ and 3 $\mu s$, and in total account for as much as 20\% of the LHC orbit.  

For ``no collision'' condition we utilize information from the beam position and timing monitors (BPTX) which are positioned 175 m around the LHC ring either side of the CMS interaction region, and produce a signal when an LHC bunch passes the monitor.  The L1 trigger requires then that neither BPTX indicates a passing bunch.   

We will record stopped particle search data at a time of no collisions in CMS.  
Consequently, the only significant physics background source 
are cosmic rays and instrumental noise.  We have used pre-collision data recorded by CMS in Summer 2009 (CRAFT, the``Cosmic Run at Four Tesla'') to measure both the cosmic and instrumental backgrounds.  CRAFT data provides an ideal control sample since it was recorded under very 
similar conditions to those we will encounter when we search for stopped gluinos.

The L1 trigger rate obtained using CRAFT data is about 200 Hz. Further rejection is required to reduce the logging rate to an acceptable level.  The main contribution to the L1 rate is HCAL electronics noise.  The High Level Trigger path first applies filters designed to reject such noise, while maintaining good efficiency for signal.  Next, jets are reconstructed, and a threshold is applied to the leading jet energy; this is the main control of HLT output rate reducing it down to 5 Hz.

To reduce the instrumental background further to acceptable levels we employ both topological 
and timing based cuts offline.  We restrict our search to jets with  $|\eta| <  1.3$.   
Since most gluino bound states are produced centrally 
due to their high mass, the endcaps and forward calorimeters do not contribute much to the signal rate. 

Studies using CRAFT data show that time and spatial profiles for the HCAL instrumental noise are 
very different from ones produced by real energy deposits. 
We make several cuts on the time profile of HCAL signal to discriminate
this noise. This finally suppresses background rate down to 0.00039~Hz.  
The total signal efficiency for those selections, estimated for $m_{\tilde{g}}$ = 300 GeV, $M_{\tilde{\chi}}$ = 50 GeV, is 
16.4\% of all stopped particles, or 52\% of all triggered events.

The measured background rate from CRAFT together with estimated signal rate from our production and decay simulations are combined as input to the last step of our simulation, and allow us to quantify our signal significance over the background by performing a large number of pseudo-experiments.  

The production cross-section for 300~GeV gluinos at $\sqrt{s} = 10$~TeV is 0.5 nb.  We consider an instantaneous luminosity scenario of $10^{32} cm^{-2} s^{-1}$, and scan gluino lifetimes from 1 $\mu s$ to 1 week.  We assume the LHC is filled with 2808 bunches per beam, and that the operating cycle consists of a 12 hour fill, followed by a 12 hour interfill period.  We also assume that the trigger is allowed to run for the entire interfill period.  

\begin{figure}
   \includegraphics[height=.2\textheight]{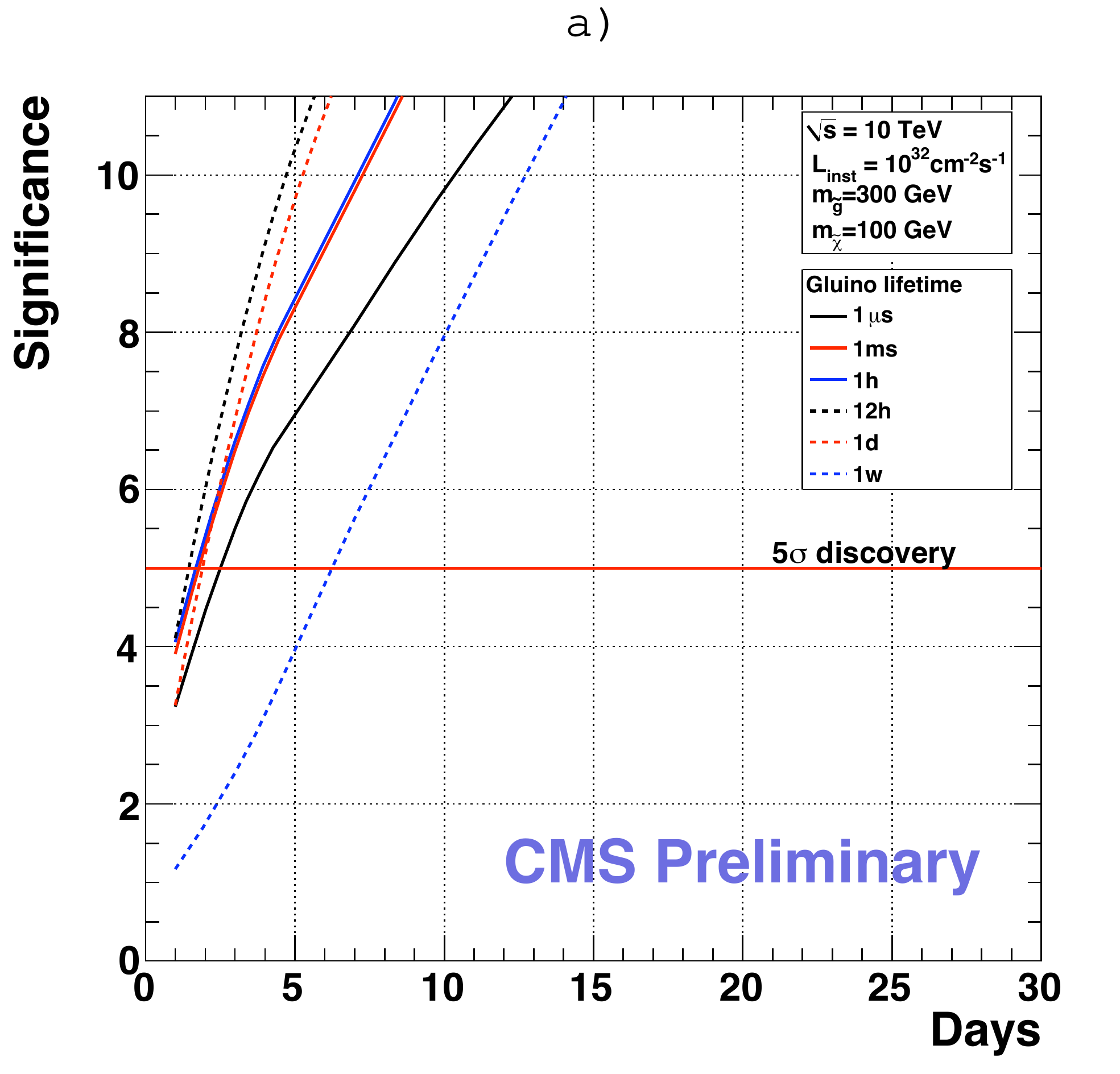}
   \includegraphics[height=.2\textheight]{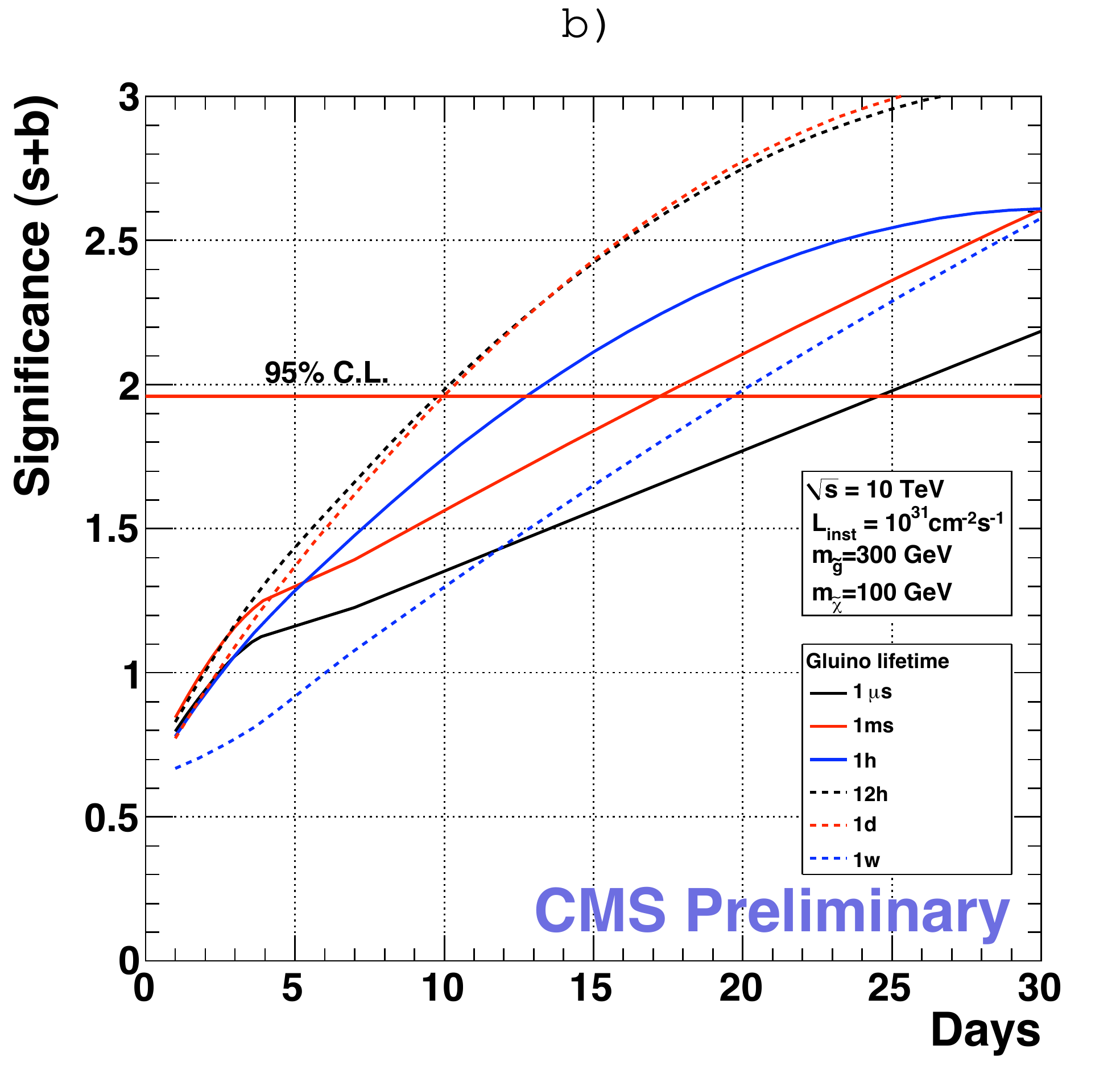}
   \includegraphics[height=.2\textheight]{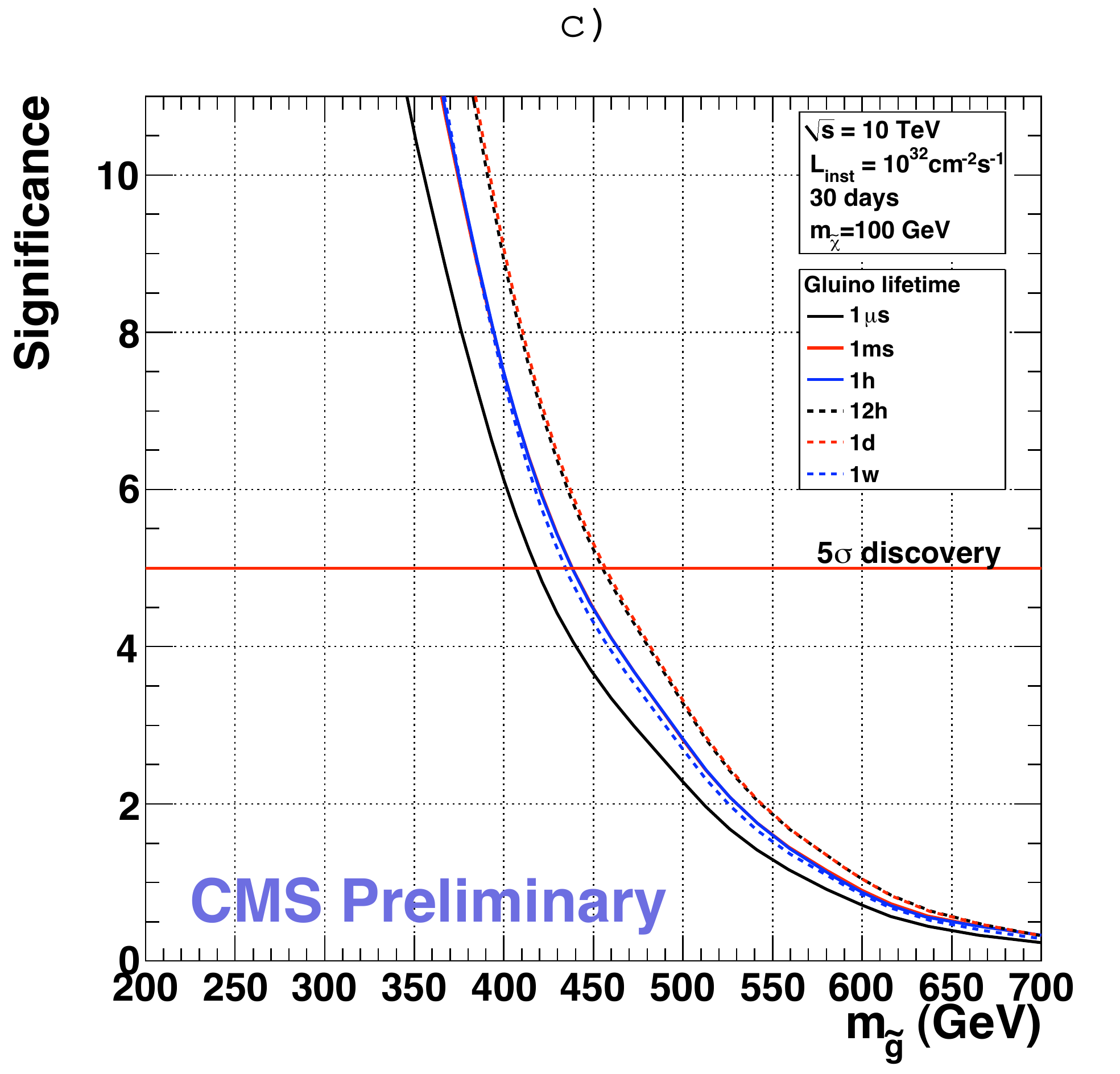}
   \caption{\label{discovery}{\bf Left:} Discoverability during both beam gaps and interfill periods in a combined counting experiment,  assuming instantaneous luminosity of $10^{32} cm^{-2} s^{-1}$. {\bf Center:} Time to reach 95\% C.L. exclusion (of signal + background hypothesis) for both beam gaps and interfill periods in a combined counting experiment, assuming instantaneous luminosity of $10^{31} cm^{-2} s^{-1}$. {\bf Right:}
Significance achievable after 30 days running, as a function of gluino mass from a counting experiment using both beamgap and interfill periods, assuming instantaneous luminosity of $10^{32} cm^{-2} s^{-1}$.  Note that the 1 ms curves are representative of a range from 10 $\mu$s to 100 s, for reasons explained in the text.}
  \end{figure}

Figure~\ref{discovery} shows the discovery reach for search during both beam gaps and interfill periods in a combined counting experiment with data obtained at instantaneous luminosity of $10^{32} cm^{-2} s^{-1}$.  The beamgap experiment is sensitive to lifetimes in the range 1 $\mu$s to hours, and the interfill experiment is sensitive to lifetimes ranging from hours to weeks.  The 1 ms curve is representative of a very large range of lifetimes, from 10 $\mu$s to 100 s, which are not shown for clarity.  In this range, the lifetime is longer than the longest gap in the LHC beam structure, yet short enough that the interfill experiment provides little live time due to the time-window.  This regime is optimal for detection, since the rate of stopped gluino decays during any gap is essentially constant.  The sensitivity to very short lifetimes will be affected by the length and number of gaps in the LHC filling scheme.

The discovery potential is much worse at an instantaneous luminosity of  $10^{31} cm^{-2} s^{-1}$.  Our backgrounds have a rate which is independent of luminosity. The expected significance after running for a given period of time goes with  $\mathcal{L}\times\sqrt{t}$, and necessary time scales quadratically with inverse luminosity. The time to reach a 95\% C.L. exclusion of the signal+background hypothesis during both beam gaps and interfill periods in low luminosity scenario is also shown in Figure~\ref{discovery}.

In addition, Figure~\ref{discovery} illustrates the significance that can be achieved after 30 days running as a function of the gluino mass, $m_{\tilde{g}}$.  The dominant factors limiting the observable range of $m_{\tilde{g}}$ is the signal cross-section.

As a summary, we have presented plans for a search for long-lived particles which have become stopped by the CMS detector.  We have described a novel calorimeter trigger that will allow us to record the subsequent decay of these stopped particles during time intervals where there are no $pp$ collisions in CMS. For models with relatively large cross-sections ($\sim 1$ nb) we find we have the potential to make a 5 sigma discovery in a matter of days with an instantaneous luminosity of $10^{32} cm^{-2} s^{-1}$.   This sensitivity improves greatly on that achieved in previous experiments~\cite{Abazov:2007ht}, in part because of the novel triggering strategy that has been implemented.

The author gratefully acknowledges the support by the Helmholtz Alliance ``Physics at the Terascale''.
This work was helped  in part by the DOE under 
contract DE-FG02-96ER40969 during the Unusual Dark Matter workshop
at the University of Oregon.

\bibliographystyle{aipproc}  
\bibliography{stopped_gluino_susy09_v2}

\end{document}